# Controllable synthesis of single-, double- and triple-walled carbon nanotubes from asphalt

Kai Xu,[a] Yongfeng Li,[a,] Chunming Xu,[a] Jinsen Gao,[a] Hongwen Liu,[b]

Haitao Yang,[b] and Pierre Richard[b]

[a]*State Key Laboratory of Heavy oil Processing, College of Chemical Engineering, China University of Petroleum, Beijing Changping 102249, China*
[b]*Beijing National Laboratory for Condensed Matter Physics, Institute of Physics, Chinese Academy of Sciences, Beijing 100190，P. R. China*

**Abstract**

Single-, double- and triple-walled carbon nanotubes (SWNTs, DWNTs and TWNTs) have been controllably synthesized by an arc discharge method using asphalt as carbon source. The morphology and structure of three kinds of products synthesized with Fe as catalyst were characterized by scanning electron microscopy (SEM), high-resolution transmission electron microscopy (HRTEM), Raman spectroscopy and energy dispersive X-ray spectroscopy (EDX). It was found that different buffer gases strongly affect the number layers in the synthesized nanotubes. In the He gas atmosphere only SWNTs were found to be synthesized by arc discharge in contrast to the case in Ar gas atmosphere in which only TWNTs were formed. In $N_2$ atmosphere both SWNTs and DWNTs were synthesized. Our findings indicate that asphalt is a suitable industrial carbon source for the growth of different type nanotubes.

[*]Corresponding author. Tel: +86-10-89739028; Fax: +86-10-89739028, E-mail: liyongfeng2004@yahoo.com.cn

# Introduction

The helical arrangement of carbon atoms on seamless of carbon nanotubes (CNTs) was first detailed analyzed in early 1991[1, 2], and after that the research in this field has grown exponentially around the world because of their unique electronic, chemical and mechanical properties, as well as potential applications in a number of fields such as in composites as reinforcing fibers, in molecular electronics as nanowires, in microscopy as probe tips, in fuel cells as cathode support and in catalysis as supports and etc. [3-6]. Depending on their structure, carbon nanotubes can be divided into two main types: single-walled nanotubes (SWNTs) and multi-walled nanotubes (MWNTs). SWNTs can be described as a sheet of graphene rolled into a seamless cylinder, with a very high length-to-diameter ratio.

Several methods have been used in the production of CNTs, such as arc discharge, laser ablation and chemical vapor deposition (CVD). In the case of laser ablation, the complexity of this technique hinders its development for large-scale production, and laser based processing in general is not suitable for scale up due to high costs of equipment and energy. Although CVD is a method which appears to present the best hope for large-scale and continuous production of CNTs owing to its relatively low cost and high yield potential [7-9], the crystal structure of CNTs is not so good and it is difficult to meet the demand of advanced electronic industry. The arc discharge method is easier to operate, more widely used, and the most important is high-quality CNTs can be synthesized in contrast to the CVD method. However, in the traditional arc discharge process, high purity graphite as carbon source is much expensive, which is used as electrode. Thus, if one cheap carbon material could be found to replace the normal graphite electrodes, CNTs would be produced in a more economical way.

For this purpose, asphalt and coal seem to be good candidates because they are cheap and abundant carbon sources in nature. Coal as carbon source for synthesis of carbon nanomaterials has been widely studied by many researchers who have ever gotten a lot of very good results [10-15]. More recently, asphalt and heavy oil residue were also used to synthesize carbon nanomaterials by the CVD method [16-22]. Prior work leaves open the question of whether the structure of CNTs from these cheap carbon sources can be further controlled. Although there are many efforts reported in the literature on the control of SWNT structure by arc discharge, such as by changing the catalyst [23], buffer gas pressure and compositions [24], shape of anode [25], less information is known about the effect of the gas type on the formation of CNTs from asphalt. In our work, it was found that the structure of CNTs produced from asphalt could be controlled by the arc discharge method. Namely, we can selectively synthesize single-, double-, and triple-walled CNTs from Chinese asphalt by arc discharge by means of controlling different buffer gas conditions. Our work is very important for CNT applications since CNT performance depends strictly on the uniformity and tunability of CNT diameter as well as the wall number.

## 2. Experimental

2.1. Preparation of carbon electrodes from asphalt

In this study, the arc vaporization method was adopted to produce various forms of CNTs. The used asphalt was from Changqing in China and three types of gases were used in this work: $N_2$ and two kinds of inert gas, He and Ar. To make an asphalt-derived electrode, the asphalt sample was crushed and sieved to 100-200 mesh, and further mixed with Fe powder (100 mesh) as catalyst in the weight ratio of 1:2. Then the mixture was filled into hollow cylinder carbon rods with bottom at one side (5 mm in inside diameter, 8 mm in outside diameter, 70 mm in length and 50 mm in deep) that are self-supporting and electrically conductive. After that we put the carbon rods filled with pitch into the oven at 100℃, and the mixture formed a paste since the softening point of asphalt was about 45℃. After cooling down to room temperature, the pitch and catalyst again formed the solid-like material which strongly bonded with the outer wall of carbon rods. In our case no cap was needed to close the hollow rods. Some composition and property data of asphalt can be found in our previous work [16].

2.2. Production and examination of CNTs

The CNTs were respectively prepared by electrically arcing carbon rods in high-purity (99.999%) $N_2$, He and Ar atmosphere in a stainless steel arc discharge chamber with an inner diameter of 165 mm and a height of 410 mm. The anode was an asphalt-derived carbon rod, the cathode was a high-purity graphite electrode (20 mm in diameter, 38 mm in length). In order to facilitate the collection of CNTs and improve the purity of the CNT samples, we placed a wire net on the top of the two electrodes in the chamber, and the distance between the wire net and the electrodes was about 5cm, which was similar to the case mentioned previously [26]. The buffer gas pressure was controlled in the 0.04–0.05 MPa range during the arc discharge experiment. The DC voltage and current for arcing were controlled at 18-20 V and 60-80 A, respectively. The distance between the two electrodes was maintained at about 1-3 mm by manually advancing the anode that was consumed during the experiment.

The weight of the graphite cathode remained the same before and after the arcing experiments. During the arc discharge process, the pitch outflow to the open end of carbon rods were immediately evaporated due to the high temperature of arc discharge. After the reaction was finished, we could not find any asphalt in the reactor. The weight of each original hollow carbon rod, the filled carbon rod and the carbon rod left after reaction are about 5.0g, 7.1g and 4.8g, respectively. Only very small amount of hollow carbon rods were evaporated, namely, the main consumed carbon source was pitch for each kind of CNT growth. We collected the film-like samples from the wire net and examined them using scanning electron microscope (SEM, FEIQuanta200F), high-resolution transmission electron microscopy (HRTEM, FEIF20), energy dispersive X-ray spectroscopy (EDX) coupled with HRTEM and Raman spectroscopy (514.6 nm laser, Horiba Jobin Yvon T64000).

## 3. Results and discussion

The macroscopic images of film-like samples synthesized respectively from asphalt in three types of gas environment during arc discharge are shown in Fig. 1. According to the figures, it is found that the film-like samples were very thin and nearly transparent. The yield of film-like sample synthesized in the He and $N_2$ conditions, as seen in Figs. 1(a) and (b), was higher than in the Ar conditions (Fig. 1(c)). In addition to the film-like deposits, fiber-like materials that deposited on the anode were also found during the arc discharge in $N_2$ atmosphere. SEM has been further employed to study the as-formed film-like deposits that were peeled off from the wire net, which was used for collecting CNTs. Figs. 2(a)-(c) show three typical SEM images of film-like samples obtained from asphalt with Fe as catalyst under He, N$_2$ and Ar atmosphere, respectively. The SEM images revealed that all the asphalt-derived film-like deposits were cotton-like threads or bundles, consisting of high-density CNTs. In addition, the deposited materials on the anode were also confirmed to be the CNTs with bundle-like morphology, as seen in Fig. 2(d).

The asphalt-derived CNTs were further examined by HRTEM. The typical HRTEM images showed in Fig. 3 exhibit various morphologies of CNTs. The low-magnification HRTEM image indicated that all the film-like materials prepared in the He conditions were composed of nanotube bundles and catalyst particles, as seen in Fig. 3(a). The high-resolution HRTEM image of Fig. 3(a) indicated that all the CNTs were SWNTs, and the diameter of an individual SWNT is 1.1 nm, as seen in Fig. 3(b). In contrast to the case of He atmosphere, the CNTs made in N$_2$ atmosphere have different morphologies, as seen in Figs. 3(c) and (d), where the bundle's diameter of SWNTs was thicker than that prepared in the He conditions shown in Fig. 3(a) and Fig. 3(c). This difference shows that the synthesized SWNTs tend to form larger bundles in N$_2$ atmosphere, which is possibly due to the formation of larger Fe catalyst nanoparticles during arc discharge process, as seen in Fig. 3 (c), compared with the case in He atmosphere. In addition, the diameter of SWNTs (Fig. 3(d)) is larger than that observed in Fig. 3(b), suggesting that the buffer gas affects the features of prepared SWNTs. More evidence of gas effect in the growth of CNTs was seen in Figs. 3(e) and (f) where most of CNTs grown in Ar atmosphere having diameter abound 4-5 nm are observed. The most striking finding is that most synthesized CNTs are triple-walled carbon nanotubes (TWNTs), as seen in Fig. 3(f). Compared with the film-like sample collected from the wire net, there are many double-walled CNTs (DWNTs) grown on the anode electrode in N$_2$ conditions during the arc discharge, as seen in Figs. 3 (g) and (h). The above results demonstrate strongly that the structure of CNTs from asphalt can be controlled by selecting the buffer gas.

Another powerful technique for studying and identifying various forms of CNTs is Raman spectroscopy, which can provide important insights into the structure of nanotubes. Figs. 4(a)-(c) show the typical Raman spectra of the as-prepared film-like CNTs obtained in the He, $N_2$ and Ar conditions, respectively. Fig. 4(a) shows the Raman spectra of the SWNTs synthesized in the He condition and a peak in the Raman spectrum can be clearly seen at 1300 cm$^{-1}$, corresponding to the D-band, which is related to a disordered or defective graphitic structure. The main peak called the G-band appears at 1578 cm$^{-1}$, which is one of the characteristic Raman peaks of the carbonaceous products with high crystallinity [27]. It is known that the ratio of the peak intensities of the G and D bands is a good index of the quality of SWNTs, and the G/D ratio of about 14 found here indicates that the synthesized nanotubes have high crystallinity. In the wave number region below 250 cm$^{-1}$, a clear peak for radial breathing modes (RBM) is observed, showing strong evidence for the formation of SWNTs. As shown in the inset, the RBM peak consists of several strong peaks at 189, 212, and 250 cm$^{-1}$. It has been well established that the frequency of the RBM is inversely proportional to the diameter of the SWNTs, and this can be correlated by the following equation, $d=234/(\omega-10)$ [28], where $\omega$ is the RBM frequency in cm$^{-1}$ and d is the SWNTs diameter in nm. According to this equation, the RBM frequencies of 189, 212, and 250 cm$^{-1}$ correspond to 1.31, 1.16, and 0.96 nm, respectively. This means that the diameters of the asphalt-derived SWNTs prepared in the He conditions are in a range between 0.96-1.31 nm. In contrast, it is found that the SWNTs grown in $N_2$ condition have a larger diameter distribution, from 1.31 to 1.65 nm by the RBM frequency, as seen in Fig.4 (b), which is in agreement with the HRTEM characterization in Figs. 3(b) and (d). The average G/D ratio of 113 implies that the quality of SWNTs synthesized in $N_2$ atmosphere is higher than that that made in He atmosphere. Also, this result demonstrates evidently that high-quality SWNTs can be synthesized from asphalt by the arc discharge method.

Fig. 4(c) shows typical Raman spectra of the TWNTs synthesized in the Ar condition, and typical RBM peaks in a low frequency region are clearly observed. The RBM frequencies of 143, 152, 168, 176 and 188 $cm^{-1}$ correspond to 1.76, 1.65, 1.48, 1.41 and 1.31 nm, respectively. According to previous work [29], the detected signal may correspond to the inner tube diameter of TWNT samples since some TWNTs have inner tube diameter of around 1.7 nm according to the HRTEM observation. A Raman spectrum of the as-grown DWNTs collected from anode electrode in $N_2$ conditions by arc discharge is shown in Fig. 4(d). Two distinct RBM peaks at low frequency of 130 $cm^{-1}$ and 198 $cm^{-1}$ are observed, corresponding to the nanotubes with diameters of 1.95 nm and 1.24 nm, respectively, and the difference is just twice the distance between graphite layers, which provides evidence that the sample is DWNTs according to the related previous report [30].

The above Raman and HRTEM characterizations have confirmed that different kinds of CNTs can be produced in different gas atmosphere during arc discharge. In addition, the elements of each kind CNTs have been measured by EDX *in situ* HRTEM observation. The element analysis of different CNTs are the same (supporting information) in which elements C and Fe origin from CNTs and the Fe catalyst, and Cu element is attributed to Cu grid for HRTEM. S element is considered to be from asphalt, which demonstrates directly that the asphalt is the carbon source for CNT formation. Therefore, our work demonstrates that SWNTs, DWNTs and TWNTs can be controllably synthesized by an arc discharge method using asphalt as carbon source. It is well known that the growth of CNTs in arc discharge is related to the density and velocity of carbon atoms, ions and clusters, temperature, and electric field in the growth region. In different gases, the plasma parameters such as the ionization degree, electron energy and electron density are different. The value of the He ionization energy (24.5 eV) is larger compared to the cases of Ar (15.76eV) and $N_2$ (15.6 eV) [31]. Therefore, the carbon plasma temperature in Ar and $N_2$ is lower than the case in the He atmosphere. Consequently, the anode asphalt evaporation ratio in Ar and $N_2$ is slower during arcing compared with the case of He atmosphere. On the other hand, the size of Fe catalyst clusters in higher arc temperature region in He atmosphere is smaller than the case of Ar and $N_2$ atmosphere, which may be attributed to the CNTs with different structures induced by different types of buffer gas during arc discharge.

## 4. Conclusion

High-quality SWNTs, DWNTs and TWNTs have been selectively synthesized for the first time by an arc discharge method with Fe as catalyst using asphalt as carbon source. In the He gas atmosphere only SWNTs were found to be synthesized compared with the case in Ar gas atmosphere in which only TWNTs were formed. In contrast to SWNTs and TWNTs, DWNTs can be synthesized on the anode electrode in $N_2$ condition. The morphology and structure of each kind of CNTs were confirmed by SEM, TEM, Raman spectroscopy and EDX. This technique opens up the possibility for economically preparing CNTs with controlled structure.

## Acknowledgements

We gratefully thank for the National Natural Science Foundation of China (Nos. 21106184 and 11274362), the Science Foundation Research Funds Provided to New Recruitments of China University of Petroleum, Beijing (No.YJRC-2011-18), Foundation for the Author of National Excellent Doctoral Dissertation of PR China (No. 201252) and Thousand Talents Program.

**Figure captions**

Figure 1 A macroscopic image of SWNTs synthesized in He atmosphere (a), SWNTs synthesized in $N_2$ atmosphere (b), TWNTs synthesized in Ar atmosphere (c).

Figure 2 SEM images of the as-grown samples for the SWNTs grown by arc discharge in He atmosphere (a), SWNTs grown in $N_2$ atmosphere (b), TWNTs grown in Ar atmosphere (c) and DWNTs grown in $N_2$ atmosphere (d).

Figure 3 Low- and high-resolution HRTEM images of the as-grown samples for the SWNTs grown by arc discharge in He atmosphere (a) and (b), SWNTs grown in $N_2$ atmosphere (c) and (d), TWNTs grown in Ar atmosphere (e) and (f) and DWNTs grown in $N_2$ atmosphere (g) and (h).

Figure 4 Raman spectra of the as-grown samples for the SWNTs grown by arc discharge in He atmosphere (a), SWNTs grown in $N_2$ atmosphere (b), TWNTs grown in Ar atmosphere (c) and DWNTs grown in $N_2$ atmosphere (d).

Figure 1

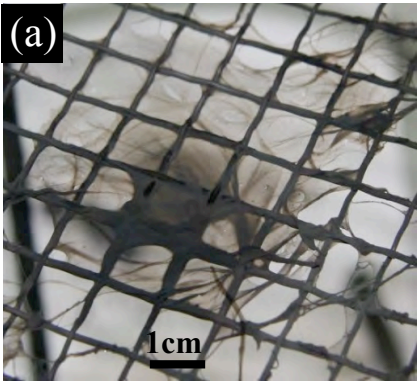 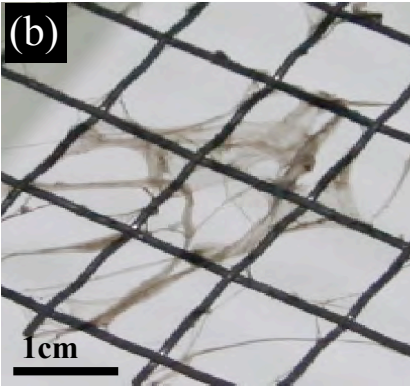

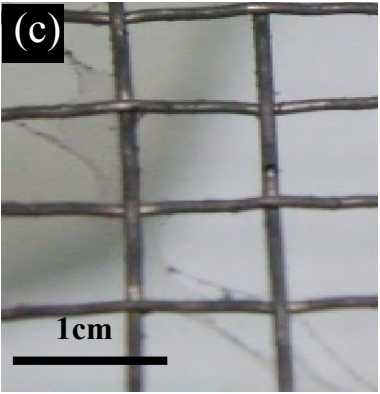

Figure2

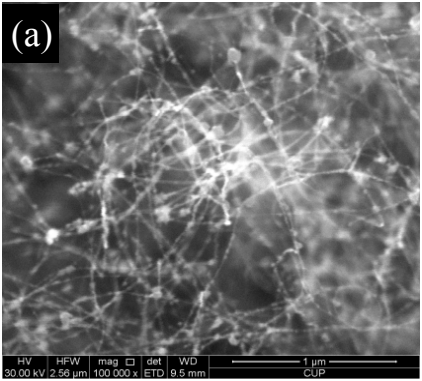 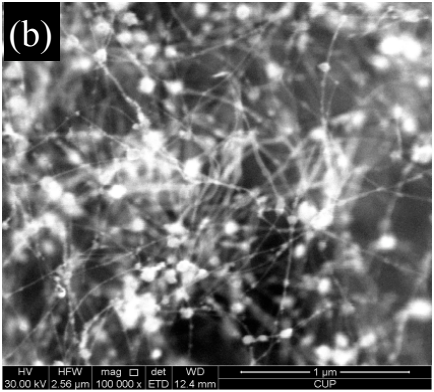
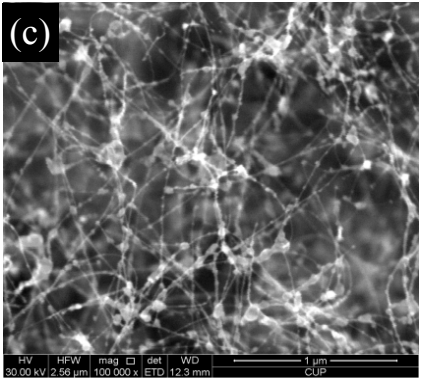 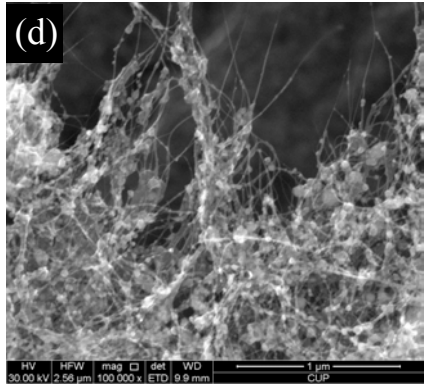

**Figure 3**

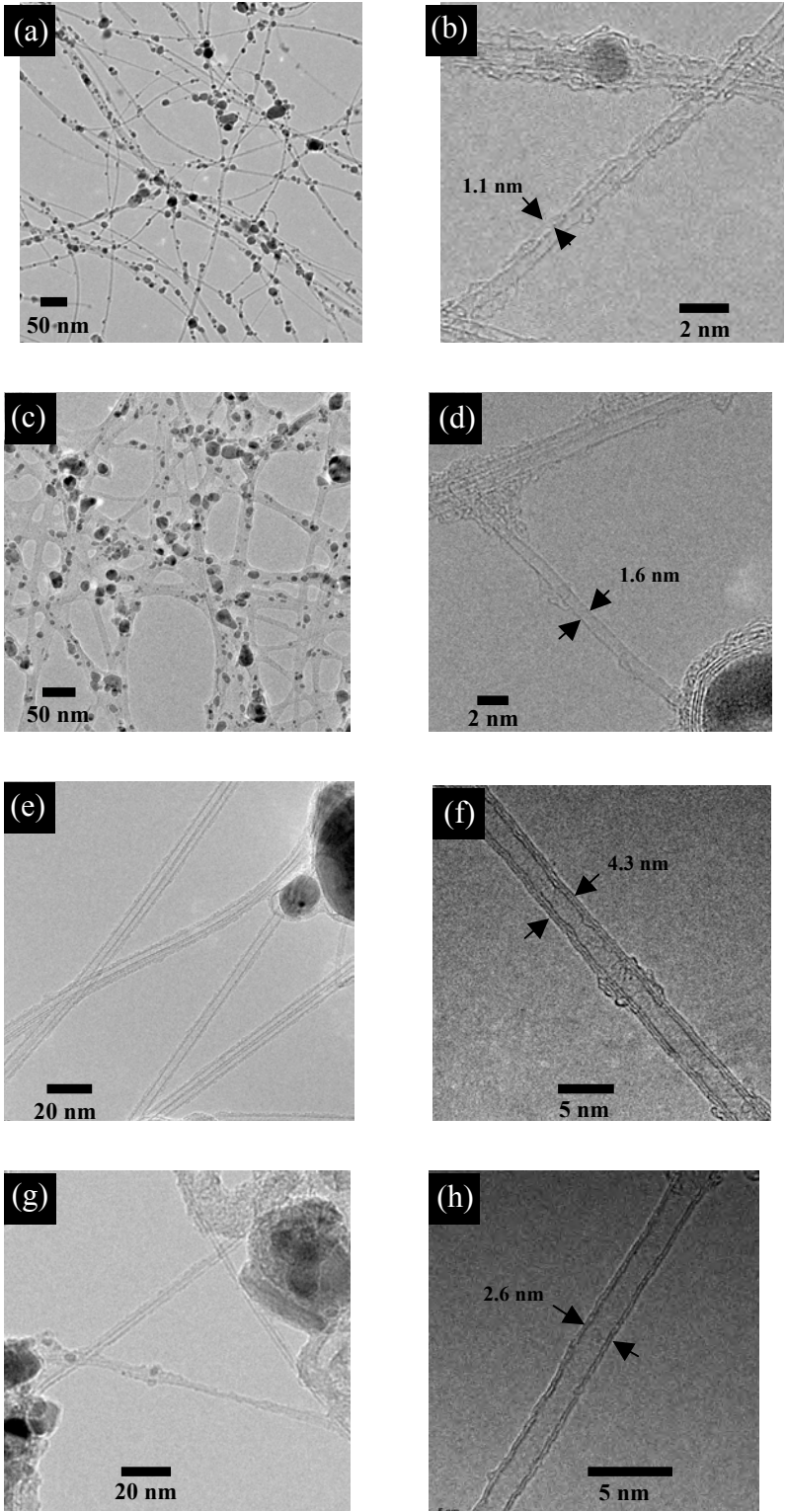

Figure 4

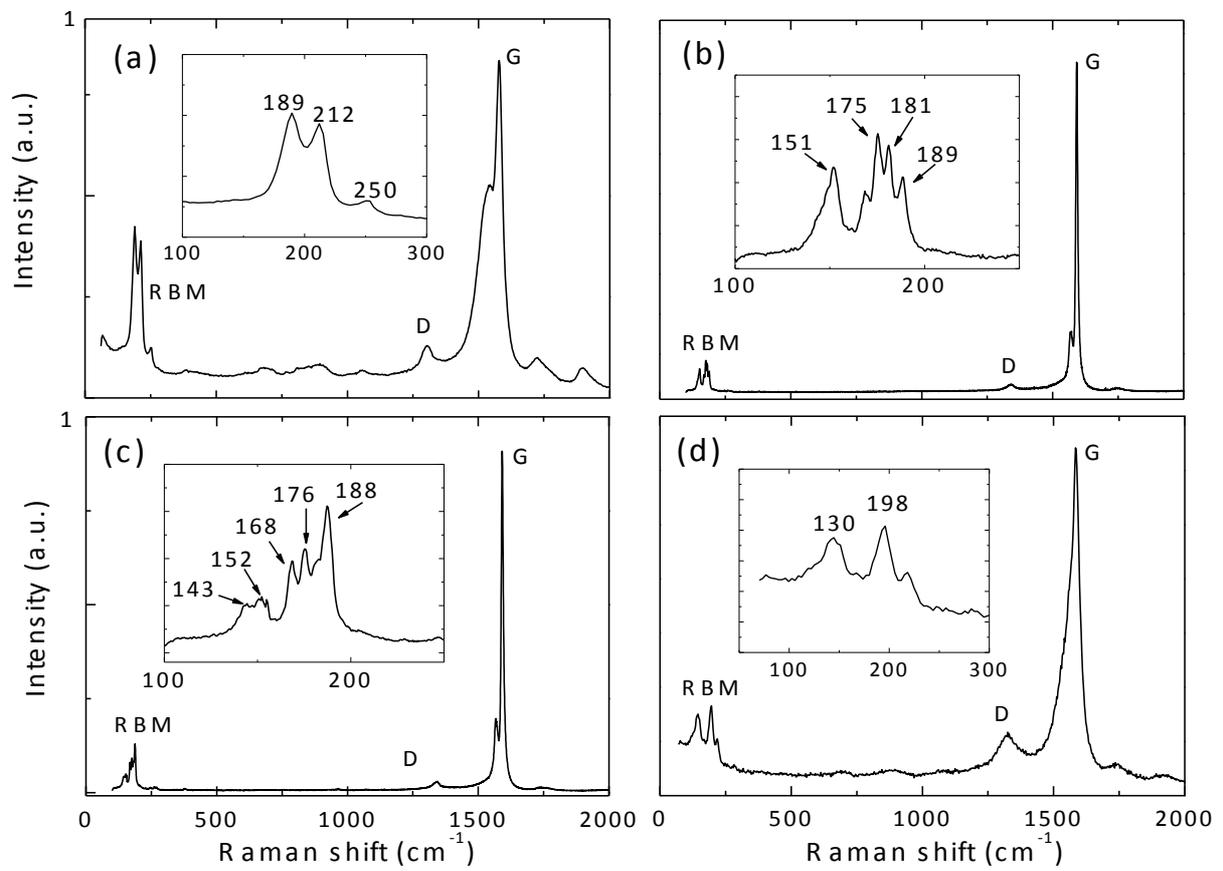